\documentclass{bioinfo}
\copyrightyear{2006}
\pubyear{2006}
\application

\newcommand{\etal}{{\it et~al}.}
\newcommand{\eg}{e.g.}
\newcommand{\ie}{i.e.}
\usepackage[normalem]{ulem}

\begin{document}
\firstpage{1}

%
\title[Overlapping modules in biological networks]{CFinder: Locating cliques and overlapping modules in biological networks} 
\author[Adamcsek \textit{et~al}]{
  Bal\'azs Adamcsek$^{\rm 1}$, 
  Gergely Palla$^{\rm 2}$,
  Ill\'es J. Farkas$^{\rm 2}$,
  Imre Der\'enyi$^{\rm 1}$, and
  Tam\'as Vicsek$^{\rm 1,2,}$\footnote{To whom 
    correspondence should be addressed.}}
\address{$^{\rm 1}$Department of Biological Physics, E\"otv\"os
University,
$^{\rm 2}$Biological Physics Research Group 
of the Hungarian Academy of Sciences,
P\'azm\'any P. stny. 1A, H-1117 Budapest, Hungary}
\maketitle

\begin{abstract}

\section{Summary:}
Most cellular tasks are performed not by individual
proteins, but by groups of functionally associated proteins,
often referred to as modules.
In a protein assocation network
modules appear as groups of 
densely interconnected nodes,
also called communities or clusters.
These modules often
overlap with each other 
and form a network of their own,
in which nodes (links) represent the modules (overlaps).
We introduce CFinder, 
a fast program locating and visualizing
overlapping, densely interconnected groups of nodes
in undirected graphs,
and allowing the user to 
easily navigate between 
the original
graph and
the web of these groups.
We show that in gene (protein) association networks
CFinder can be used to predict the function(s) 
of a single protein and
to discover 
novel modules.
CFinder is also 
very efficient 
for locating the cliques of large sparse graphs.

\section{Availability:}
CFinder (for Windows, Linux, and Macintosh)
and its manual can be downloaded from
{\color{blue} \uline{http://angel.elte.hu/clustering}}.

\section{Contact:} cfinder@angel.elte.hu
\end{abstract}

\footnotesize

\section{Introduction}

High-throughput experimental techniques,
\eg, protein-protein interaction (PPI) 
and mRNA expression methods,
have largely advanced our knowledge 
about the functioning of the cell.
{\it Gene (protein) association networks} 
integrate the broadest possible set of evidence
-- including high-throughput data --
on protein linkages:
they provide an integrated 
{\it list of binary interactions} 
(\citealp{STRING,DIP})
and allow the discovery of
previously uncharacterised cellular systems 
(\citealp{DateMarcotteNBT}).
One major goal of current research efforts
is to elucidate 
how the observed behaviours of an entire cell
can be understood in terms of the interactions
of its protein modules.
To identify such modules, a common approach is 
to search for groups of densely interconnected nodes 
in the cell's protein association network
(\citealp{Bader,Rives}).
Note, however, that modules strongly overlap.
According to the CYGD database (\citealp{CYGD}),
in {\it Saccharomyces cerevisiae} 
the number of proteins 
in known protein complexes 
(modules where the participating proteins 
physically interact at the same time)
vs. the sum of the sizes of these complexes
is $2750/8932$. 
Thus, most protein modules probably share many of their 
proteins with other modules.

\begin{figure}[]
\centerline{\includegraphics[width=0.45\textwidth]{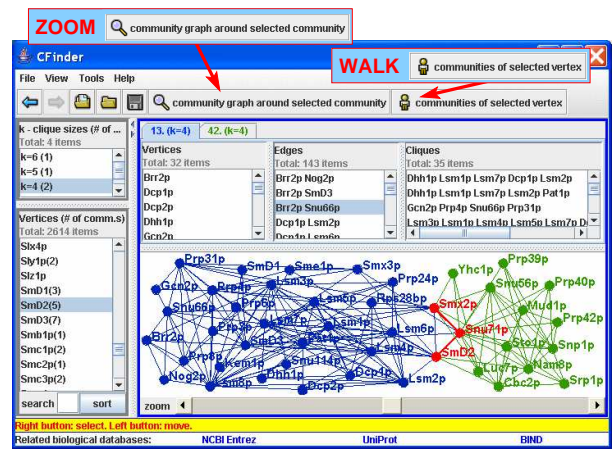}}
\caption{(colour online) Modules of the protein 
SmD2 in the 
DIP (Database of Interacting Proteins) 
``yeast core'' data set
as shown by the {\it Vertices} view of CFinder.
The two modules are coloured 
blue and green. Overlaps, 
\ie, proteins and links participating in more than one
module are red.
Enlarged on the top 
are two special buttons enabling the navigation between
the original network (a part of which is displayed)
and the web of its modules.}
\label{fig:screenshots}
\end{figure}

We introduce CFinder, a platform-independent, stand-alone  
application locating overlapping groups of densely interconnected nodes in
graphs, and illustrate its use on the network of gene
associations in the yeast genome.
We decided to maintain CFinder as an independent program 
(as opposed to a package plugin),
because it can be employed by potential users belonging to diverse fields
including, in addition to bioinformatics, economics or sociology.

Generic graph visualisation and analysis programs
(\citealp{pajek})
are frequently used for the layout and structural analysis of
networks.
Recent bioinformatics software platforms
(\citealp{cytoscape}), 
on the other hand, 
enable the user to 
integrate many different types of data,
\eg, PPI,
expression levels, 
and annotation information.
CFinder reads a list of binary interactions,
performs a search for 
dense subgraphs (groups),
and -- unlike several
currently used algorithms
(\citealp{NewmanEPJB}) 
--
it allows for any node to belong 
to more than one group.
Due to its algorithm and implementation,
CFinder is efficient 
for networks with millions of nodes
and, as a byproduct of its search,
the full clique overlap matrix of the network is determined.
Below we will show that
in gene association networks
CFinder's results can be used to predict
novel modules and novel individual 
protein functions.

\section{Overview of CFinder}

\begin{figure*}[!tpb]
\centerline{\includegraphics[width=1.0\textwidth]{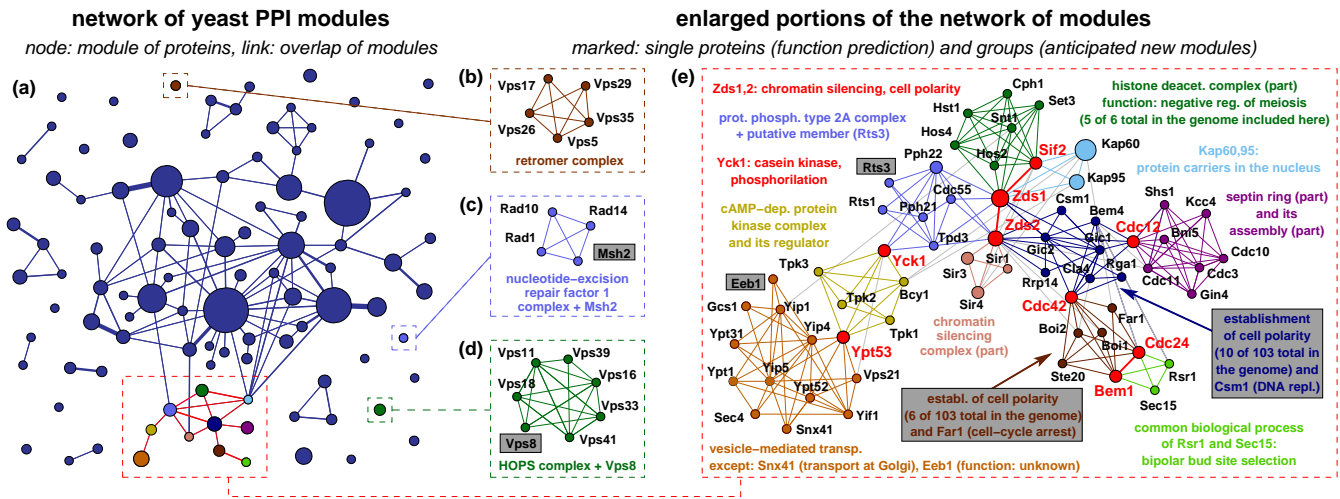}}
\caption{(colour online) {\bf (a)} 
The network of modules mapped by CFinder in the DIP
``yeast full'' data set ($k=4$).
{\bf (b-d)}
In addition to locating known complexes, 
CFinder often groups together a known complex 
with one additional protein,
allowing the improvement 
of the functional annotation of that protein (Msh2, Vps8).
{\bf (e)}
Zooming into the network of modules and adding 
Gene Ontology (GO) annotation terms 
(i) produces a {\it detailed and well-structured layout} 
of the original network of proteins,
(ii) provides {\it characterisation for individual proteins} (Eeb1, Rts3) 
and 
(iii) {\it predicts new modules} 
(dark blue and brown, see text). 
} 
\label{fig:commNW}
\end{figure*}

The {\it input of CFinder} is a file containing
strings and numbers ordered into three columns; 
in each row the first two
strings correspond to the 
two end points of a link and the third item is the
weight of this link.

The computational core of CFinder was implemented in C++, 
while the visualisation and analysis components 
were written in Java. 
The {\it search algorithm} uses the
Clique Percolation Method (CPM, see \citealp{Derenyi})
to locate the {\it $k$-clique percolation clusters} 
of the network 
that we interpret as modules.
A $k$-clique is a 
complete subgraph on
$k$ nodes ($k=3,4,\ldots$),
and two $k$-cliques are said to be adjacent, 
if they share exactly $k-1$ nodes.
A $k$-clique percolation cluster 
consists of
(i) all nodes that can be reached 
via chains of adjacent $k$-cliques from each other
and (ii) the links in these cliques.
Note that larger values of $k$
correspond to a higher stringency 
during the identification of dense groups
and provide smaller groups
with a higher density of links inside them.
For both local and global analyses in a network,
we suggest 
using such a value of $k$
(typically between $4$ and $6$)
that provides the user with the richest
group structure 
(see \citealp{Palla}).
In the presence of link weights
CFinder can apply lower and upper cutoff values to keep only 
the set of connections judged to be significant by the user.

The {\it user interface} 
of CFinder offers several views of the analysed
network and its module structure. 
As an example,
Fig.~1 shows the modules of the protein Pwp2
in the DIP ``yeast core'' network
(\citealp{DIP}) at clique size $k=4$.
Alternative views currently available in CFinder are 
``Communities'' (displaying the identified modules),
``Cliques'',
``Stats'' 
(statistics of, \eg, module and overlap sizes)
and ``Graph of communities''. 
The special buttons ``forward'', ``back'', ``zoom''
and ``walk'' 
allow a quick navigation between the views.
A wide variety of visualisation 
settings can be adjusted
in the ``Tools'' menu.

Figure~2 displays the network of modules produced by CFinder
($k=4$) in the DIP ``yeast full'' data set.
In the complete map (a)
each node represents a module,
the area of a node is proportional 
to the number of proteins in the corresponding module,
and the width of a link is proportional to 
the number of proteins shared by the two modules.
Panel (b) shows a
previously known complex identified by CFinder.
Panels (c) and (d) both display a known complex 
grouped together 
with one additional protein (Msh2 and Vps8, respectively),
leading to an improved functional annotation
of that protein.
In panel (e)
Eeb1 (function currently unknown) 
is grouped together 
with proteins participating 
in vesicle-mediated transport, 
thus, 
{\it we predict this to be a key function of Eeb1}. 
Proteins in 
the marked dark blue and brown 
groups 
of panel (e)
cooperate 
on the establishment of cell polarity, 
a function performed by a total of $103$ proteins in the cell. 
(Please, see colour figure online.)
We anticipate that these two groups are 
{\it biologically meaningful, novel modules} 
within that larger set of $103$ proteins. 
[Gene names and annotations were handled with Perl tools, 
\eg, GO::TermFinder (\citealp{GOtermfinder}).]

\section{Acknowledgements}

The authors acknowledge funding from 
the Hungarian Sci. Res. Fund, 
OTKA (Grants No. D048422 , F047203, and T049674).

\end{document}